\documentstyle[aps,prl,twocolumn]{revtex}

\def\be{\begin{equation}}
\def\ee{\end{equation}}

\newcommand{\bra}[1]{\mbox{$\langle #1 |$}}
\newcommand{\ket}[1]{\mbox{$| #1 \rangle$}}
\newcommand{\braket}[2]{\mbox{$\langle #1  | #2 \rangle$}}
\newcommand{\proj}[1]{\ket{#1}\!\bra{#1}}

\newtheorem{defi}{Definition}
\newtheorem{teo}{Theorem}
\newtheorem{lemma}{Lemma}

\begin{document}

\draft

\title{Entanglement monotones}

\author{Guifr\'e Vidal}

\address{e-mail:guifre@ecm.ub.es \\
Departament d'Estructura i Constituents de la Mat\`eria\\ 
Universitat de Barcelona\\ Diagonal 647, E-08028 Barcelona, Spain}

\date{\today}

\maketitle

\begin{abstract}
 In the context of quantifying entanglement we study those functions of a multipartite state which do not increase under the set of local transformations. A mathematical characterization of these monotone magnitudes is presented. They are then related to optimal strategies of conversion of shared states. More detailed results are presented for pure states of bipartite systems. It is show that more than one measure are required simultaneously in order to quantify completely the non-local resources contained in a bipartite pure state, while examining how this fact does not hold in the so-called asymptotic limit. Finally, monotonicity under local transformations is proposed as the only natural requirement for measures of entanglement.

\end{abstract}


\bigskip 

\section{Introduction and main results}
\label{seccio1}

Entanglement is an essential ingredient of quantum information. Indeed quantum correlations play a crucial role in most applications of this blooming field, since they are behind several processes which would not be possible in a purely classical context. Thus, in addition to being of interest from a fundamental point of view, entanglement is a most praised quantum resource.

Entanglement lacks, however, a complete quantification. Given an arbitrary state it is not known yet, for instance, what processes requiring non-local resources can be accomplished, nor which its cost of preparation in terms of other states is. This information would allow for an optimal use of the resources contained in a shared state, and should arguably be provided by a proper quantification of entanglement.

 Local transformations are very much related to the quantification of entanglement as a resource. As a typical framework one encounters several classically communicating parties which share a composite system in an entangled state, on which they can only operate locally. The non-local resources of the shared state are known to degrade under the local action of the parties. Then, among all possible local transformations, optimal strategies of conversion are concerned with the minimization of the degradation of entanglement when a given shared state is converted into another. And thus, through a deeper knowledge of local transformations we expect to acquire a sufficient theoretical command on entanglement, so that it can be transformed, and hence exploited, optimally. 

 A remarkable step forward in this direction was achieved with the introduction of the entropy of entanglement \cite{BBPS}, a measure of the entanglement for pure states of bipartite systems. The entropy of entanglement quantifies, asymptotically, the non-local resources of a large number of copies of a pure state.
 Several extensions of this measure to mixed states have been also proposed, such as entanglement of formation and distillable entanglement \cite{BdVSW} and relative entropy of entanglement \cite{VedrPl}. They share the aim of quantifying the entanglement of an infinite number of copies of a mixed state.

 However, as far as a finite number of shared states is concerned, the quantification of entanglement is still in early stages. Already in \cite{BBPS} difficulties were reported when attempting to find a local conversion strategy involving just one copy of an entangled pure state which could achieve the asymptotic ratios of conversion which follow from the entropy of entanglement. Later work \cite{LoPop} proved that as a matter of fact the optimal ratio at which the entanglement of a finite number of copies can be concentrated by local means is often lower than the asymptotic one, thus showing that one can not rely on asymptotic results when quantifying entanglement in the finite scenario. Additional progress in this direction has been reported recently in \cite{Nie,Vid,Har,JonPl}  and in \cite{KLM} regarding pure and mixed states respectively.

 This work is intended to contribute to a better understanding of the entanglement of a multipartite state. It focuses on those magnitudes which behave monotonically under local transformations of the system. Such state-dependent magnitudes, which will be referred as entanglement monotones, will be argued to play a significant role in the study of entanglement. The main results we will present are:

\begin{itemize}
\item a complete characterization of entanglement monotones in terms of elementary local transformations, together with a justification of their importance in relation to entanglement transformations (sec. \ref{seccio2});

\item a tool for building infinitely many entanglement monotones for bipartite systems. Our recipe shows how to build {\em all} possible entanglement monotones for pure states of these systems (sec. \ref{seccio3});

\item an exhaustive classification of pure states of bipartite system into local equivalence classes, which follows from the study of a family of entanglement monotones, the $\alpha$-entropies of entanglement, and which shows that more than one magnitude is required in order to characterize the entanglement of pure states (sec. \ref{seccio4}-\ref{seccio5}); 

\item an analysis of monotonicity in the so-called asymptotic regime. We show how to conciliate the fact that many entanglement monotones exist with the uniqueness of entanglement measures in the {\em asymptotic limit}\cite{PopRo} (sec. \ref{seccio4}).

\end{itemize}

 Finally, several proposals have been made about the requirements any reasonable measure of entanglement should fulfill (see, for instance \cite{VedrPl,PopRo,VPRK,VT}). It will be proposed in sec.\ref{seccio5} to diminish the list of a priori requirements simply to monotonicity under local transformations. From this point of view any entanglement monotone could be regarded as a measure of entanglement.

\section{Local transformations and entanglement monotones}
\label{seccio2}

 In this section we will characterize those state-dependent magnitudes which have a monotone behavior under local transformations of a multipartite system. By local transformation we understand any transformation that the state $\rho$ of the system may undergo when the parties sharing it are allowed to perform local quantum operations assisted with classical communication (LQCC). Notice that if $\mu(\rho)$ is a non-decreasing monotone, then obviously $-\mu(\rho)$ is a non-increasing monotone. Consequently we can restrict our considerations to non-increasing monotones.

\begin{defi}
We call entanglement monotone EM any magnitude $\mu(\rho)$ that does not increase, on average, under local transformations. 
\end{defi}

 We will start with a review of the elementary steps any local transformation can be decomposed into. This will allow us to establish a set of necessary and sufficient conditions for a magnitude $\mu$ to be an EM.  Then we will relate monotonicity to entanglement transformations.

\subsection{Local quantum operations and classical communication.}

 Let us thus consider several parties, denoted by $i=A, B, ...$, sharing a composite system ${\cal Q}$. We will assume that the Hilbert space ${\cal H} ={\cal H}_A \otimes {\cal H}_B \otimes ...$ of the degrees of freedom of ${\cal Q}$ which are relevant concerning entanglement is finite dimensional, so that the state of these degrees of freedom is a density matrix $\rho:{\cal H}\longrightarrow {\cal H}$ of finite rank \cite{finit}. We will also assume that the parties, which are allowed to communicate unrestrictedly between them, can operate on their subsystems only locally, but without any further limitation within the observance of quantum theory rules. Present technological restrictions will therefore not be taken into account.

 In order to characterize EMs we need a precise description of the set of local transformations a multipartite state $\rho$ can undergo. We will focus our attention on the two distinct ingredients of each such transformations, namely local quantum operations LQs and classical communication CC, separately. Notice that LQs correspond to the most general quantum operation the parties can perform on their subsystems, each party acting independently of what the rest of them do, and they are therefore genuinely local, whereas CC allows for the parties to correlate their local actions. Thus due to CC classical correlations between the subsystems can be introduced or enhanced.

 Let us analyze first LQs. What is the effect of a unilocal quantum operation in a multipartite state? A compact answer can be presented using the formalism of completely positive superoperators \cite{Kraus}. The most general quantum operation ${\cal E}_{i,k}$ performed locally by only party $i$, say party A for clearness, is implemented by means of a set of product operators  $\tilde{O}_{k,j} = A_{k,j}\otimes I_B \otimes ...$ with the restriction $\sum_{k,j} A_{k,j}^{\dagger}A_{k,j} \leq I_A$, where $A_{k,j}:{\cal H}_A \longrightarrow {\cal H}_A'$ \cite{canvi}, the index $k$ refers to the distinct outcomes if a measurement takes part in this local operation, and $I_l$ is the identity operator in ${\cal H}_l$. On the other hand any such superoperators corresponds to some unilocal quantum operation. As a result of this operation the state of the composite system becomes,
\be
\rho_k = \frac{{\cal E}_{i,k}(\rho)}{p_k} =\frac{1}{p_k} \sum_j \tilde{O}_{k,j}\rho \tilde{O}_{k,j}^{\dagger}
\label{localtransf}
\ee
with probability $p_k = $Tr$[{\cal E}_{i,k}(\rho)]$ ($i=A$ in our example). We can succinctly write this transformation as

\vspace{4mm}

L1) {\em Unilocal quantum operation}:
\begin{equation}
 \rho\longrightarrow \{p_k, \rho_k\}, 
\label{unilo}
\end{equation}
where $\rho_k$ and $p_k$ are related to the initial state $\rho$ of the multipartite system by means of a unilocal, completely positive superoperator ${\cal E}_{i,k}$ as in eq. (\ref{localtransf}).

\vspace{4mm}

There is a remark to be made about transformation L1 concerning transmission of information between the parties. In writing transformation L1 we have implicitly adopted the point of view of the party which performs the quantum operation. In absence of CC the rest of the parties, being ignorant of the actual outcome of the operation (but knowing just which is the operation that has been performed), will have to describe the transformed state as $\rho'= \sum_k p_k \rho_k$ (cf. eq. (\ref{unilo})). We adopt here, in order to avoid such proliferation of inequivalent descriptions, the point of view of an external observer to whom the parties explain (via a one-way classical channel) what the results of their operations are \cite{external}. Then, from our external observer's point of view, the most general LQ ---i.e. such that several, isolated parties intervene successively--- corresponds simply to composing several unilocal quantum operations as described in eq. (\ref{unilo}).

For later analysis we have found it convenient to further decompose any unilocal quantum operation L1 into four basic steps, each one corresponding to a simple physical manipulation. Thus any LQ can be achieved by composing the following, conceptually simpler quantum operations performed locally by one of the parties only:

\vspace{4mm}

L1.1) {\em Unilocal unitary transformation}:
\begin{equation}
\rho \longrightarrow \rho' =\tilde{U} \rho \tilde{U}^{\dagger}
\end{equation}
where $\tilde{U}$ is a unitary operator such that it is the identity for all parties but for one, for which it is a unitary transformation. For instance, in a bipartite system unilocal unitaries are implemented by $\tilde{U}=U_A\otimes I_B$ or $\tilde{U}=I_A\otimes U_B$, where $U_i$ is a unitary in the local subsystem of party $i$.

\vspace{2mm}

L1.2) {\em Unilocal von Neumann measurement}:
\begin{equation}
 \rho\longrightarrow \{\rho_k, p_k\},
\end{equation}
 which follows from just one of the parties performing a von Neumann measurement, not necessarily complete, on its local subsystem, $\rho$ becoming the state $\rho_k$ with probability $p_k$. Such a measurement transforms the state $\rho$ by means of projectors that are the direct product of the identity for all the parties but one, for which it is a projector (not necessarily of rank one). For instance if, again in a bipartite system, party A measures, the projectors that implement such a unilocal measurement are of the form $P_k = P_{A,k}\otimes I_B$, $\{P_{A,k}\}$ being orthogonal and realizing the identity operator $I_A$ .
\vspace{2mm}

L1.3) {\em Addition of} an uncorrelated {\em ancilla} $\tilde{\cal Q}$:
\begin{equation}
\rho\longrightarrow \rho' = \rho\otimes\rho_{\tilde{\cal Q}},
\label{addition}
\end{equation}
where $\rho_{\tilde{\cal Q}}$ is the state of the ancilla, added by just one party to its subsystem.
\vspace{2mm}

L1.4)  {\em Dismissal of a local part} $\tilde{\cal Q}$ of the whole system:
\begin{equation}
 \rho\longrightarrow \rho' = Tr_{\tilde{\cal Q}}[\rho],
\end{equation}
where $Tr_{\tilde{\cal Q}}[.]$ indicates partial trace over $\tilde{\cal Q}$. $\tilde{\cal Q}$ being local, it is originally held by only one of the parties.
\vspace{2mm}

 Let us notice here that L1.1 and L1.2 refer to LQs that the parties perform on the system, which stays the same, but whose state is modified, whereas L1.3 and L1.4 consist essentially of an actual change in the number of components of the composite system under consideration. Notice also that while L1.1 and L1.3 can always be reversed by local means, both L1.2 and L1.4 are typically irreversible.

 Let us move to consider the role of classical communication in local transformations. In a multi-party local strategy consisting of a series of LQs, CC allows the parties to make some of these LQs dependent on the result of some previous ones, when the latter are of stochastic nature (that is, when they involve, e.g., a measurement and thus more than one outcome is possible a priori).
 This allows for local transformations which are not in the set of LQs. And thus, for instance, entanglement transformation strategies such as entanglement distillation can be made more powerful when based on LQCC than when based only on LQs \cite{BdVSW}.

 Characterizing mathematically the whole set of local transformations proves to be more involved than when no CC is allowed between the parties. However, if such coordinated local actions are divided again into basic steps, we happen to simply recover transformations L1.1-L1.4 (that is, L1) as the only constituent pieces that involve a proper interaction with the multipartite system. Accordingly, the basic local transformation which we will consider additionally, and which comes from allowing CC between the parties, does not involve any further manipulation of the system. We introduce it next by means of a concrete example.

 Two parties,  Alice and Bob, share a bipartite system ${\cal H} ={\cal H}_A \otimes {\cal H}_B$ consisting of the spin of two pairs of identical particles. Let us denote by $\varrho_1$ the state of one pair, of which both Alice and Bob possess each one particle, and by $\varrho_2$ the state of the other pair, which is also shared \cite{explanation}, the state of the whole system being thus $\rho=\varrho_1\otimes\varrho_2$. 
We will consider two versions of a correlated {\em stochastic} transformation, which simply consists in the parties dismissing one of their pairs of particles. Suppose Alice uses a {\em random} process to decide which of her two particles she is to throw away. Then she can tell Bob of her intentions, so that both of them can dismiss a particle corresponding to the same pair. Suppose also that the a priori probabilities, $q_1$ and $q_2 (\equiv 1- q_1)$, that the parties end up sharing $\varrho_1$ and $\varrho_2$ respectively are known to them. We will write down this stochastic, local but classically correlated transformation as
\be
\rho \longrightarrow \left\{ \begin{array}{c} 
    q_1, \varrho_1 \\
    q_2, \varrho_2
   \end{array} \right.
\label{ambinfo}
\ee
After this transformation the parties share either state $\varrho_1$ or state $\varrho_2$, and they could now further act on the system depending on which of the two states they actually share. Consider, alternatively, a second version of the previous transformation in which no record of the pair of particles being dismissed is kept \cite{exemple}. Then the transformation reads
\be
\rho \longrightarrow  \rho' = q_1\varrho_1 + q_2 \varrho_2,
\label{senseinfo}
\ee
where $\rho'$ collects all the information which is now available (that is, the possible outcomes $\varrho_k$ and corresponding probabilities $q_k$ in the first version of the transformation). 

 The only difference between the final state in transformations (\ref{ambinfo}) and (\ref{senseinfo}) is in the amount of information the parties have about the system, i.e., in the information which is {\em available} to them \cite{info}. Notice that this information is highly relevant for the parties if they regard the state they share as a physical resource. Notice also that transformation (\ref{senseinfo}) can be obtained from the output of transformation (\ref{ambinfo}) if the parties a posteriori dismiss or lose the information of which pair has been kept. Since dismissal of available information can be done {\em locally}, we will consider as local also the transformation  
\be
\left.
 \begin{array}{c} 
    q_1, \varrho_1 \\
    q_2, \varrho_2
   \end{array} \right\} \longrightarrow \rho' = q_1\varrho_1 + q_2 \varrho_2,
\label{canviinfo}
\ee 
which has to be understood in a statistical sense over many repetitions of the dismissal of information at the end of transformation (\ref{ambinfo}). 

 Consequently when studying monotonicity under LQCC, we will also consider a suitable generalization of the transformation (\ref{canviinfo}):

\vspace{2mm}

L2)  {\em Decrease in the information} available about the state of the system:
\be
 \{ q_k, \varrho_k\}\longrightarrow \rho' = \sum_{k} q_k \varrho_k,
\label{L2}
\ee
where  $\{ q_k, \varrho_k\}$ is {\em any} ensemble realizing $\rho'$.
\vspace{2mm}

\subsection{Local monotonicity.}

 From transformations L1 and L2 we obtain the following two necessary and sufficient conditions for a magnitude $\mu$ to be, on average, non-increasing under local transformations, and thus an entanglement monotone:

\vspace{2mm}

C1) For any $\rho$ and any unilocal quantum operation ${\cal E}_{i,k}$ performed by any party $i$,
\be
\mu(\rho) \geq \sum_k p_k \mu(\rho_k),
\label{C1}
\ee
where (cf. L1)
\begin{eqnarray}
p_k &=& \mbox{Tr}[{\cal E}_{i,k}(\rho)] \nonumber \\
\rho_k &=& \frac{1}{p_k} {\cal E}_{i,k}(\rho). \nonumber 
\end{eqnarray}
\vspace{2mm}

C2) For any ensemble $\{q_k,\varrho_k\}$
\be
\hspace{4mm} \sum_{k} q_k \mu(\varrho_k) \geq \mu(\rho)
\label{C2}
\ee
where $\rho = \sum_{k}q_k \varrho_k$ (cf. L2).

\vspace{2mm}

 Conditions C1 and C2 are necessary for monotonicity because L1 and L2 correspond to local transformations. In order to see that they are also sufficient, it is useful to regard any local transformation as a series of classically correlated unilocal quantum operations, in some of which information about the actual result of a measurement or about the actual performance of a stochastic operation is not made available. Condition C1 assures monotonicity, on average, under any unilocal quantum operation of the series individually. That these LQs can be dependent on other operations is completely irrelevant as far as monotonicity of $\mu$ is concerned. Finally C2 assures that monotonicity is preserved if some of the information about actual performances and results is not made available during the local transformation.
      
 Let us make some more remarks on these two conditions. Firstly, they imply that $\mu$ takes a constant value $\mu_0$ for separable states \cite{separable}, which is its minimal one. This is so because separable states can be reversibly converted into each other by means of LQCC, that is, all separable states are locally equivalent, and because the parties can always make the system end up in a separable state. This value being irrelevant, we will always take $\mu_0 = 0$ (by considering $\mu'\equiv\mu - \mu_0$ instead of $\mu$ if necessary), so that without loss of generality we will consider only non-negative definite EMs which vanish for separable states, i.e.
\be
\mu(\rho) \geq 0; ~~~~~ \rho_s \mbox{ separable } \Rightarrow \mu(\rho_s) = 0.
\label{pos}
\ee
Similarly, if $\mu$ is an EM, so is $\mu'\equiv a \mu$ for any $a\geq 0$, so that we could normalize this quantity to take a given value for a suitable state of reference.

Our second remark concerns the statistical character of monotonicity in C1 and C2. Notice that if $\rho_k$ is one of the possible final states of a local generalized measurement as in eq. (\ref{C1}), it may happen that $\mu(\rho_k) > \mu(\rho)$, that is, a concrete performance of a LQ may result in an increase of $\mu$. C1 implies, however, that this measurement can not be used to increases the magnitude $\mu$ systematically, since on average (over many performances of the measurement) $\mu$ will be at most conserved. Similarly, it may happen that $\mu(\varrho_k) < \mu(\rho)$ for some $\varrho_k$ in eq. (\ref{C2}), so that $\mu$ may occasionally increase as the parties {\em forget} information, although C2 assures again that they can not obtain such an increase systematically.

 Thirdly, condition C1 is equivalent to the following four independent conditions:
\vspace{2mm}

C1.1) $\hspace{4mm}\mu(\rho) = \mu(\tilde{U}\rho \tilde{U}^{\dagger})\hspace{4mm}$ (cf. L1.1),
\vspace{2mm}

C1.2) $\hspace{4mm} \mu(\rho) \geq \sum_{k} p_k \mu(\rho_k)$, \hspace{2mm} if $\{\rho_k, p_k\}$ describes the state resulting from a (not necessarily complete) unilocal von Neumann measurement on $\rho$ (cf. L1.2), 
\vspace{2mm}

C1.3) $\hspace{4mm} \mu(\rho) = \mu(\rho\otimes\rho_{\tilde{\cal Q}}),\hspace{2mm}$ $\rho_{\tilde{\cal Q}}$ being any state of a local ancilla $\tilde{\cal Q}$ (cf. L1.3), and
\vspace{2mm}

C1.4) $\hspace{4mm} \mu(\rho) \geq \mu(Tr_{\tilde{\cal Q}}[\rho])$, $\hspace{2mm}$for any local part $\tilde{\cal Q}$ of the system (cf. L1.4).

Notice that monotonicity has been replaced with invariance in C1.1 and C1.3 due to the local reversibility of transformations L1.1 and L1.3.
 
The set of conditions C1.1-C1.4 are very transparent in terms of basic unilocal quantum operations (L1.1-L1.4) and we will need them for later reference. Nevertheless we will use condition C1 in most of the rest of this work because of its compactness.

\subsection{Optimal local conversion strategies and monotonicity.}
\label{util}

 We move now to explore in what sense EMs may be useful when studying entanglement transformations. For this purpose we consider a local strategy (i.e. a local transformation), aiming at converting an initial shared state $\rho$ into a final state $\tilde{\rho}$. Such a strategy may involve local measurements, on whose result the success of the conversion depends. Then the probability $p(\rho\rightarrow \tilde{\rho})$ that such a local strategy lead to a successful conversion satisfies
\be
 p(\rho\rightarrow\tilde{\rho}) \leq \frac{\mu(\rho)}{\mu(\tilde{\rho})}
\label{inequal}
\ee
for any EM $\mu$.
 Indeed, if $\{\rho_k\}$ denote the possible final states of the local strategy ($\rho_1 \equiv \tilde{\rho}$) and $p_k$ denotes the probability of obtaining $\rho_k$ ($p_1 \equiv p(\rho\rightarrow\tilde{\rho})$), then the monotonicity of $\mu$, together with eq. (\ref{pos}), implies that
\be
\mu(\rho) \geq \sum_k p_k \mu(\rho_k) \geq p_1 \mu(\rho_1),
\ee
from which eq. (\ref{inequal}) follows immediately. We can, however, go further and announce
 
\begin{teo} 
 The maximal probability of success $ P(\rho \rightarrow \tilde{\rho}) $ in the conversion of $\rho$ into $\tilde{\rho}$ that the parties can achieve by means of a strategy involving only LQCC is given by
\be
P(\rho\rightarrow\tilde{\rho}) = \min_{\mu} \frac{\mu(\rho)}{\mu(\tilde{\rho})},
\label{useful}
\ee
where the minimization is to be performed over the set of all EMs (rescaled as to satisfy eq. (\ref{pos})).
\end{teo}
{\bf Proof:} it suffices to see that, as it follows from its definition, $P(\rho\rightarrow\tilde{\rho})$ is an EM as a function of $\rho$, and that, obviously, $P(\tilde{\rho}\rightarrow\tilde{\rho})=1$, so that $\mu(\rho) = P(\rho\rightarrow\tilde{\rho})$ can be used on the RHS of eq. (\ref{useful}).$\Box$

 Eqs. (\ref{inequal}) and (\ref{useful}) reveal an intimate relation between entanglement transformations and EMs. Eq. (\ref{inequal}) shows that we can use EMs to test the optimality of a given local conversion strategy. Indeed, if there is an EM $\mu$ which saturates eq. (\ref{inequal}), then the strategy is optimal. And, on its turn, eq.(\ref{useful}) indicates that during any conversion strategy which is known to be optimal at least one EM is conserved. 

 The study of monotonicity, however, proves to be even more fruitful. Through EMs we get an insight into the nature of entanglement itself, as we will show in subsequent sections. In particular, the domain of equivalence under LQCC of pure states of a bipartite system will follow from the study of a family of EMs, namely the $\alpha$-entropies of entanglement. Such classes of local equivalence suggest that one can not do with just one measure of entanglement to quantify the non-local resources of shared states. This and the mere existence of many EMs for pure states are in apparent contradiction with well established asymptotic results \cite{PopRo}. Once more, by studying how EMs behave in the asymptotic limit we will see that a consistent picture emerges.
 
\section{Building entanglement monotones for pure states}
\label{seccio3}

 The following sections focus on entanglement monotones for pure states of systems which are shared by two parties. We will begin their study by showing that EMs are in one to one relation with unitarily invariant, concave real functions on the space of density matrices over ${\cal C}^n$ (up to trivial rescaling of these functions). 

It follows from Theorem 1 that the maximal probability $P(\rho\rightarrow\tilde{\rho})$ of success in a local conversion $\rho\rightarrow\tilde{\rho}$ can be obtained by performing a minimization over the whole set of EMs as in eq. (\ref{useful}). However, apart from some monotonic measures of entanglement so far proposed \cite{llista}, little is known about this set. The next result is useful for building EMs for any state, be it pure or mixed, of bipartite systems. Moreover we will use it to build the whole set of EMs for pure states $\psi$ of these systems, from which $P(\psi \rightarrow \tilde{\psi})$ could be in principle obtained \cite{probabilitat}.

 Given any real function $f:{\cal T}({\cal C}^{n}) \longrightarrow {\cal R}$ on the space ${\cal T}({\cal C}^{n})$ of density matrices $\sigma$ on ${\cal C}^n$ such that it is 
\begin{itemize}
\item invariant under the transformation of $\sigma$ by any unitary matrix $U:{\cal C}^{n}\longrightarrow{\cal C}^{n}$ , i.e.
\be
f(U\sigma U^{\dagger}) = f(\sigma), ~~~~ \mbox{and}
\label{A1}
\ee
\item concave, i.e.
\be
f(\sigma) \geq \lambda f(\sigma_1) + (1-\lambda)f(\sigma_2), 
\label{A2}
\ee
for any $\lambda \in [0, 1]$ and any $\sigma_1,~\sigma_2 \in {\cal T}({\cal C}^{n})$ such that $\sigma = \lambda \sigma_1 + (1-\lambda)\sigma_2$,
\end{itemize}
a (non-increasing) EM $\nu:{\cal T}({\cal C}^{n}\otimes{\cal C}^{n})\longrightarrow {\cal R}$ can be derived by defining it for pure states (normalized vectors $\ket{\psi}\in{\cal C}^{n}\otimes{\cal C}^{n}$) as
\be
\nu(\psi) \equiv f(\mbox{Tr}_B[\proj{\psi}]) ~~~(~=~f(\mbox{Tr}_A[\proj{\psi}]) ~),
\label{A3}
\ee
where Tr$_B[.]$ is the partial trace over subsystem $B$, and by extending it over the whole set of density matrices ${\cal T}({\cal C}^{n}\otimes{\cal C}^{n})$ as
\be
\nu(\rho) \equiv \min_{\Upsilon_{\rho}} \sum_j p_j \nu(\psi_j),
\label{A4}
\ee
where the minimization is to be performed over all the pure-state ensembles $\Upsilon_{\rho}=\{p_j,~\psi_j\}$ realizing $\rho$, i.e., such that $\rho=\sum_j p_j\proj{\psi_j}$. Normalization $\nu_0 = 0$ (cf. eq. (\ref{pos})) corresponds then to requiring $f(\sigma) = 0$ for pure states $\sigma = \proj{\phi}$.

 In order to prove that $\nu(\rho)$ satisfying eq. (\ref{A1})-(\ref{A4}) is an EM it suffices to check that it fulfills conditions C1 and C2 in the context of two parties, namely Alice and Bob. We will next examine only the case when Bob acts locally on his subsystem, since by construction $\nu$ is symmetric under exchange of parties.  

\begin{teo}
 Any function $\nu(\rho)$ satisfying eqs. (\ref{A1})-(\ref{A4}) is an EM, i.e.,
\be
\mbox{eqs. (\ref{A1})-(\ref{A4})} ~~\Longrightarrow~~ C1,C2.
\ee
\label{tool}
\end{teo}
{\bf Proof:} Let us assume first that $\rho$ in condition C1 is a pure state $\psi$. Recall that if Bob performs a quantum operation $\{{\cal E}_{B,k}\}$ on his subsystem (where $k$ labels different outcomes if at some stage of his local manipulations Bob performs a measurement), then the parties will obtain the state $\rho_k \equiv \frac{1}{p_k}{\cal E}_{B,k}(\proj{\psi})$ with probability $p_k\equiv $Tr$[{\cal E}_{B,k}(\proj{\psi})]$. Since the reduced density matrix seen by Alice cannot be affected by Bob's local action alone, as this would imply an arbitrarily fast transmission of information from Bob to Alice, $\sigma\equiv$ Tr$_B[\proj{\psi}]$ equals the ensemble average of the reduced density matrices $\sigma_k\equiv$ Tr$_B[\rho_k]$, i.e.
\be
\sigma = \sum_k p_k\sigma_k.
\label{una}
\ee 
For each $k$, let $\{r_{kl},\psi_{kl}\}$ be a pure-state ensemble realizing $\rho_k$ optimally, in the sense that
\be
\nu(\rho_k) = \sum_l r_{kl} \nu(\psi_{kl}),
\label{dues}
\ee
where $r_{kl} > 0$, $\sum_l r_{kl} = 1$ and $\rho_k = \sum_l r_{kl} \proj{\psi_{kl}}$, and define $\sigma_{kl} \equiv $Tr$_B [\proj{\psi_{kl}}]$. Then eqs. (\ref{una}),  (\ref{A2}) and (\ref{dues}) successively imply:
\begin{eqnarray}
\nu(\psi) \equiv f(\sigma) = f(\sum_k p_k\sigma_k) = f(\sum_{kl} p_kr_{kl} \sigma_{kl}) \geq \nonumber \\ 
\sum_{kl} p_kr_{kl} f(\sigma_{kl}) \equiv \sum_{kl} p_kr_{kl} \nu(\psi_{kl}) = \sum_k p_k\nu(\rho_k),
\label{purs}
\end{eqnarray}
so that C1 is satisfied.

 Suppose now that $\rho$ in C1 is any mixed state. We need introduce the following states and optimal pure-state ensembles OPSE (as in eq.(\ref{dues})):

\begin{enumerate}
\item Let $\rho_k \equiv \frac{1}{p_k}{\cal E}_{B,k}(\rho)$ ($p_k\equiv$Tr$[{\cal E}_{B,k}(\rho)]$) be the final state when transforming $\rho$ according to ${\cal E}_{B,k}$ as in C1.
\item Let $\{t_j,\eta_j\}$ be an OPSE of $\rho$.
\item For each $j$, let $\rho_{jk} \equiv \frac{1}{t_{jk}} {\cal E}_{B,k}(\eta_j)$ ($t_{jk}\equiv$ Tr $[{\cal E}_{B,k}(\eta_j)]$) be the final state when transforming $\eta_j$ according to ${\cal E}_{B,k}$. Eq. (\ref{purs}) implies that
\be
\nu(\eta_j) \geq \sum_k t_{jk} \nu(\rho_{jk}),
\label{tres}
\ee
while linearity of ${\cal E}_{B,k}$ implies that
\be
\rho_k = \frac{1}{p_k} \sum_j t_j t_{jk} \rho_{jk}.
\label{quatre}
\ee
\item For each pair $j,k$ let $\{t_{jkl},\psi_{jkl}\}$ be an OPSE of $\rho_{jk}$.
\end{enumerate}
Then, using eq. (\ref{tres}) we obtain
\begin{eqnarray}
\nu(\rho) = \sum_j t_j \nu(\eta_j) \geq \sum_{jk} t_j t_{jk} \nu(\rho_{jk}) = \nonumber \\
\sum_{jkl} t_j t_{jk} t_{jkl} \nu(\psi_{jkl}) \geq \sum_k p_k\nu(\rho_k),
\end{eqnarray}
where in the last inequality we have also used that 
\be
\nu(\rho_k) \leq \sum_{jl} \frac{t_jt_{jk}t_{jkl}}{p_k} \nu(\psi_{jkl}),
\ee
which follows from  $\rho_k = \frac{1}{p_k}\sum_{jl} t_jt_{jk}t_{jkl} \proj{\psi_{jkl}}$ (cf. eq. (\ref{quatre})) and from eq. (\ref{A4}).

 In order to prove C2 we just need to consider an OPSE $\{q_{kl},\phi_{kl}\}$ for each $\varrho_k$ in C2. Since then $\rho = \sum_{kl} q_k q_{kl}\proj{\phi_{kl}}$ , eq. (\ref{A4}) implies that
\be
\nu(\rho) \leq \sum_{k}q_k\sum_lq_{kl} \nu(\phi_{kl})\equiv\sum_k q_k \nu(\varrho_k).\Box
\ee

 Eq. (\ref{A1}) means that $f(\sigma)$ is a function of the eigenvalues of $\sigma$ only. These eigenvalues correspond, through eq. (\ref{A3}), to the square of the coefficients $\{\sqrt{\alpha_i}\}$ in the Schmidt decomposition
\be 
\ket{\psi} = \sum_{i=1}^n \sqrt{\alpha_i} \ket{i_A\otimes i_B}
\label{Schmidt}
\ee
 of the pure state $\psi$. Therefore any $f(\sigma)$ in eqs. (\ref{A1}) and (\ref{A2}) is determined by a concave, symmetric function $g_f: \Delta_n \longrightarrow R$ defined on the $n$-dimensional simplex $\Delta_n$ \cite{simplex}:
\be
f(\sigma) = g_f(\{\alpha_i\}). 
\label{funciog}
\ee 

 As examples of EMs built using Theorem \ref{tool} consider, for {\em any} function $\hat{f}(x)$ concave in the interval $x\in[0,1]$, the function $\hat{f}:{\cal T}({\cal C}^n) \longrightarrow {\cal T}({\cal C}^n)$ defined to act, with $\sigma=\sum_{i=1}^n \alpha_i \proj{i_A}$ the spectral decomposition of $\sigma$, as $\hat{f}(\sigma) \equiv \sum_{i=1}^n \hat{f}(\alpha_i) \proj{i_A}$. Then $f(\sigma) \equiv $Tr$[\hat{f}(\sigma)]$ satisfies eqs. (\ref{A1}) and (\ref{A2}) (see \cite{We} for concavity). Note that in this case $g_f(\{\alpha_i\}) = \sum_{i=1}^n \hat{f}(\alpha_i)$ in eq. (\ref{funciog}). Then the entropy of entanglement corresponds to taking $\hat{f}(x) = -x\log_2x$.

 Notice that C1.1 implies that {\em any} EM $\mu$ on a bipartite system ${\cal C}^{n}\otimes{\cal C}^{n}$ depends, when evaluated on a pure state $\psi$, only on its Schmidt coefficients $\{\sqrt{\alpha_i}\}$ or equivalently, on its partial trace Tr$_B [\proj{\psi}]$ through a unitarily invariant function. If we now show that, in addition, such a function needs to be concave, then we will have indicated, by means of eqs. (\ref{A1}), (\ref{A2}) and (\ref{A3}), how to construct any EM for pure states.

\begin{lemma} 
Any entanglement monotone $\mu$ on ${\cal C}^{n}\otimes{\cal C}^{n}$ can be written, when evaluated on pure states $\psi$, as a concave function $h(\sigma\equiv $Tr$_B [\proj{\psi}] )$ of its partial trace, that is
\be
\mu(\psi) = h(\sigma)
\label{mupurs}
\ee
where
\be
h(\sigma \equiv\frac{\sigma_1 + \sigma_2}{2}) \geq \frac{h(\sigma_1) + h(\sigma_2)}{2}
\label{concave}
\ee
for any pair of density matrices $\sigma_1, \sigma_2 \in {\cal T}({\cal C}^n)$.
\end{lemma}
{\bf Proof:} As mentioned above, eq. (\ref{mupurs}) follows from condition C1.1. In order for $h(\sigma)$ to be concave it suffices to demand that, given any $\sigma$, eq. (\ref{concave}) hold for any couple $\sigma_1$ and $\sigma_2$ in an open subset of ${\cal T}({\cal C}^n)$ that contains $\sigma$, for local concavity implies global concavity. We will now prove that monotonicity under LQCC requires, through condition C1, that $h(\sigma)$ be locally concave. More specifically, for any density matrix $\sigma$ in the interior of ${\cal T}({\cal C}^n)$ --i.e., with no vanishing eigenvalues-- corresponding to a pure state $\psi$ in ${\cal C}^{n}\otimes{\cal C}^{n}$, we will show that there is an open subset ${\cal T}_{\sigma} \in {\cal T}({\cal C}^n)$ containing $\sigma$ such that, for any pair of density matrices $\sigma_1$ and $\sigma_2 \equiv 2\sigma - \sigma_1$ $\in {\cal T}_{\sigma}$, a unilocal quantum operation ${\cal E}_{B,k}$, $k=1,2$, exists that with probability $p_k =\frac{1}{2}$ transforms $\psi$ into $\psi_k$, where $\sigma_k = $Tr$_B [\proj{\psi_k}]$. Then, if $\mu$ is to be non-increasing under such unilocal quantum operation, $h(\sigma)$ has to satisfy eq. (\ref{concave}), and therefore it has to be concave.  
 
 Indeed, for any pure state $\psi$ with Schmidt decomposition as given in eq. (\ref{Schmidt}) and corresponding $\sigma = \sum_i \alpha_i \proj{i_A}$, and for any density matrix $\sigma_1 \equiv \sigma + \delta\sigma$ ($\delta\sigma^{\dagger} = \delta\sigma$, Tr$[\delta\sigma] = 0$) whose components are constrained by $|(\delta\sigma)_{ij}| <  \min_l(\{\alpha_l^2\})$, consider the generalized local measurement on Bob's side implemented by the operators $O_1$and $O_2$ on ${\cal C}^n$, defined as 
\be
O_{1,2}  \equiv \sqrt{\frac{I_B\pm \tau}{2}},~~~~~~ O_1^{\dagger}O_1 + O_2^{\dagger}O_2 = I_B,
\label{mesura}
\ee
where $\tau$ is a self-adjoint operator whose components are $(\tau)_{ij} \equiv \frac{(\delta\sigma)_{ij}}{\sqrt{\alpha_i\alpha_j}}$ (notice that since for any two normalized $\ket{\phi}, \ket{\phi '} \in {\cal C}^n$, $|\bra{\phi} \tau \ket{\phi '}| \leq n \max(|(\tau)_{ij}|) \leq n \max(|\frac{(\delta\sigma)_{ij}}{\sqrt{\alpha_i\alpha_j}}|) < n \min(\alpha_i)\leq 1$, the operators $I_B\pm \tau$ are positive defined, as they should in eq. (\ref{mesura})). Since $\ket{\psi_k} = \sqrt{2} I_A\otimes O_k \ket{\psi}$, this is the announced unilocal quantum transformation.$\Box$

Therefore we can state the following theorem:
\begin{teo}
The restriction of any entanglement monotone on pure states $\ket{\psi} \in {\cal C}^n \otimes {\cal C}^n$ is given by a unitarily invariant, concave function of the partial trace $\sigma\equiv$Tr$_B[\proj{\psi}]$.
\label{T3}
\end{teo}

\section{Irreversibility of entanglement transformations}
\label{seccio4}

 Theorems \ref{tool} and \ref{T3} show how any EM for bipartite pure states can be built. Theorem \ref{tool} implies, in addition, that infinitely many EMs exist. In this section we use a particular family of these monotones to classify pure states into local equivalence classes. Two pure states turn out to be locally equivalent (that is, reversibly convertible into each other, with certainty and through local transformations) if, and only if, they are related by bilocal unitaries. These results are known not to hold in the asymptotic limit, where in addition only one EM exists \cite{PopRo}. We then analyze these apparently contradictory facts, to conclude that uniqueness of EMs in the asymptotic limit is fully compatible with our results.

\subsection{Local uncertainty as entanglement monotone.}

 We begin by presenting a family of EMs, the $\alpha$-entropies of entanglement. They have a straightforward physical interpretation, since they measure the uncertainty in the outcome of local measurements. These monotones are additive for pure states, and can be regarded as a generalization of the entropy of entanglement, which appears as a particular, privileged case. Their monotonicity under LQCC allows to classify pure states into local equivalence classes, whereas their behavior in the asymptotic limit is paradigmatic, in that it sheds some light on how every EM but the entropy of entanglement loses its monotonicity in this limit.

We need first a brief review of measures of uncertainty. One way of quantifying the uncertainty associated with a random variable which can take $n$ different values, each one having an a priori probability of occurrence $p_k$ --- where $\sum_{k=1}^n p_k = 1$--- is by means of the $\alpha$-entropy or information of order $\alpha$ \cite{Renyi}, defined as

\begin{equation}
S_{\alpha}(\{p_k\}) \equiv \frac{1}{1-\alpha} \log_2 \sum_{k=1}^n p_k^{\alpha}, \,\,\,\, \alpha \geq 0.
\end{equation}
 We will consider here only the case $\alpha\in [0,1]$. Recall that  $S_0(\{p_k\})$ can be defined as $\lim_{\alpha \rightarrow 0^+} S_{\alpha}$, and it is just the logarithm of the number of non-vanishing elements in $\{p_k\}$, whereas $S_1(\{p_k\})$, defined as $\lim_{\alpha \rightarrow 1^{\!-}} S_{\alpha}$, is the Shannon entropy
\begin{equation}
S_1(\{p_k\}) = - \sum_{k=1}^m p_k \log_2 p_k.
\end{equation}

 In quantum mechanics the outcome of a complete von Neumann measurement on a system with Hilbert space ${\cal H} = {\cal C}^n$ prepared in the state $\rho$ is a random variable, whose probability distribution $\{p_k\}_{k=1 \cdots n}$ depends both on the state and on the specific measurement. For any fixed $\rho$ it turns out that the minimal uncertainty, as quantified by $S_{\alpha}(\{p_k\})$, $\alpha\in [0,1]$, occurs when the set of $n$ orthogonal rank one projectors $\{P_k= \proj{k}\}$ that characterize the complete von Neumann measurement are such that $\{\ket{k}\}$ are the eigenvectors of $\rho$ (as it can be proved by making use of the concavity of the function $f(x) \equiv x^{\alpha}$, $x \geq 0$, $\alpha \in [0,1]$). Thus the ${\alpha}$-entropy of the eigenvalues of $\rho$\cite{We},
\begin{equation}
S_{\alpha}(\rho) = \frac{1}{1-\alpha} \log_2 Tr [\rho^{\alpha}], \,\,\,\,\,\, \alpha \in [0, 1], 
\end{equation}
can be regarded as a measure of the minimal uncertainty associated with $\rho$.

 Consider now a pure state $\psi$ of a bipartite system ${\cal C}^n \otimes{\cal C}^n$, and the minimal uncertainty that can be reached in a complete von Neumann measurement performed locally by one of the parties, say Alice. It turns out that only if $\psi$ is a product state $\phi_A\otimes\phi_B$ then Alice will be able to predict with certainty the outcome of her properly chosen measurement (which includes a projector onto $\phi_A$). For a general pure state the minimal local uncertainty, which we will call the $\alpha$-entropy of entanglement $E_{\alpha}$ of $\psi$, is given by
\begin{equation}
E_{\alpha}(\psi) \equiv \frac{1}{1-\alpha} \log_2 Tr [\sigma^{\alpha}], \,\,\,\, \sigma = Tr_B[\proj{\psi}],
\label{defpur}
\end{equation}
or, alternatively, in terms of the Schmidt coefficients of $\psi$ (cf. eq. \ref{Schmidt}),
\be
E_{\alpha}(\psi) = \frac{1}{1-\alpha} \log_2 \sum_{i=1}^n \alpha_i^{\alpha},
\label{coeff}
\ee
where it becomes manifest that Alice and Bob have the same minimal uncertainty, also that the ${\alpha}$-entropy of entanglement is an additive quantity, which reduces to the entropy of entanglement in the limit $\alpha \rightarrow 1$. Notice that Tr$[\sigma^{\alpha}]$ is a unitarily invariant function of $\sigma$, also that it is concave for $\alpha \in [0,1]$ (cf. comment after eq. (\ref{funciog})). Then, it follows from the fact that the logarithm is a concave, non-decreasing function and therefore preserves concavity, that the $\alpha$-entropy is a unitarily invariant, concave function of the partial trace $\sigma$. Consequently, by extending its definition to mixed states as in eq. (\ref{A4}), Theorem {\ref{tool}} can be applied to prove that the $\alpha$-entropies of entanglement are EMs.

 {\em Additivity} is a most convenient feature of $E_{\alpha}$ when it comes to relating the value of these monotones for a state $\psi$ to that of several copies $\psi^{\otimes N}$ of it, and we will take advantage of this property when analyzing the asymptotic limit. Another of their features, which will play a major role later on, concerns the {\em continuity} of $E_{\alpha}$ when regarded as a function of a suitable measure of the degree of resemblance of states. As such a measure we will use here the fidelity $F(\psi,\psi') = |\langle\psi | \psi'\rangle|^2$, this quantity being the probability that a system in state $\psi$ {\em behaves} as if it were in state $\psi'$. Notice that for any $\alpha \in (0,1]$, $E_{\alpha}$ is a continuous function of the fidelity, whereas $E_0$ is not. 
Let us illustrate what we mean with an example: consider, among the pure states of a ${\cal C}^2\otimes{\cal C}^2$ system, the family of states $\ket{\psi(\theta)} \equiv \cos\theta \,\ket{1_A\otimes 1_B} + \sin\theta \,\ket{2_A\otimes 2_B}$, $\theta \in [0, \frac{\pi}{4}]$. Their fidelity with respect to the product state $\ket{\psi(0)} =\ket{1_A\otimes 1_B}$ is $F(\theta,0) = \cos^2 \theta$, and one can write 

\begin{eqnarray}
\Delta E_{\alpha} \equiv E_{\alpha}(\psi(\theta))- E_{\alpha}(\psi(0))\nonumber \\= \frac{1}{1-\alpha} \log_2 (\cos^{2\alpha} \theta + \sin^{2\alpha} \theta)
\label{exemplea}
\end{eqnarray}
as a continuous function of $F(\theta, 0)$ for any $\alpha \in (0,1]$: 
\be
\Delta E_{\alpha} (F) = \frac{1}{1-\alpha} \log_2 (F^{\alpha} + (1-F)^{\alpha}).
\ee
 On the other hand the $\alpha = 0$ case, which is $E_0(\psi)= \log_2 m_{\psi}$,  $m_{\psi}$ being the number of non-vanishing Schmidt coefficients of $\psi$, is a discontinuous magnitude with respect to $F$. Taking up again the previous example: 
\begin{equation}
\Delta E_0(F)= \left\{ \begin{array}{cl}  0 & \mbox{if $F=0$} \\ 1 & \mbox{if $F>0$.} \end{array} \right.
\label{exempleb}
\end{equation} 

\subsection{Reversible conversion strategies and conversion ratios.}
 
 Here we would like to address the question: given any two bipartite pure states, $\psi$ and $\tilde{\psi}$, what do they have to satisfy in order for them to be convertible, in a reversible manner and with certainty, by means of LQCC? \cite{reversible}. 

The answer to this question is of interest when one regards entanglement as a non-local resource. For imagine the parties could convert, with certainty, the state $\psi$ into the state $\tilde{\psi}$, and the state $\tilde{\psi}$ into the state $\psi$ , by means of LQCC. Then these two states would be {\em equivalent} as far as the non-local resources needed to create them are concerned, also as for the non-local resources that can extracted from them.

\begin{defi}
We call the states $\psi$ and $\tilde{\psi}$ locally equivalent if, and only if,  strategies using only LQCC exist which convert each one of these states into the other with certainty. We then write $\psi \sim \tilde{\psi}$.
\end{defi}

 Any two states 
\begin{eqnarray}
\ket{\psi} &=& \sum_{i=1}^n \sqrt{\alpha_i} \ket{i_A\otimes i_B} ~~~\alpha_i\geq \alpha_{i+1}\geq 0, \\
\ket{\tilde{\psi}} &=& \sum_{i=1}^n \sqrt{\tilde{\alpha}_i} \ket{i_A'\otimes i_B'} ~~~\tilde{\alpha}_i\geq \tilde{\alpha}_{i+1}\geq 0,
\end{eqnarray}
with identical Schmidt coefficients (i.e. $\alpha_i = \tilde{\alpha}_i~~\forall i=1,\cdots,n$) are locally equivalent, since by means of local unitaries the parties can always reversibly convert, with certainty, the two local orthonormal bases $\{\ket{i_A}\}$ and $\{\ket{i_B}\}$ into $\{\ket{i_A'}\}$ and $\{\ket{i_B'}\}$ respectively. Consequently without loss of generality we can restrict our considerations to the Schmidt coefficients of $\psi$ and $\tilde{\psi}$. We then have:

\begin{teo} $\psi$ and $\tilde{\psi}$ are locally equivalent if, and only if, they are related by local unitaries, i.e.
\be
\psi \sim \tilde{\psi} ~~~~\Longleftrightarrow~~~~ \alpha_i = \tilde{\alpha_i} ~~~\forall i=1,\cdots,n.
\ee
\end{teo}
{\bf Proof:} It remains to be proved that if for any $i$, $\alpha_i \neq \tilde{\alpha}_i$, then at least one of the states cannot be converted into the other with certainty by means of LQCC. We can see this with the help of the $\alpha$-entropies of entanglement. Indeed, it can be checked that, unless $\alpha_i = \tilde{\alpha}_i$ $\forall i=1, \cdots n$,  $E_{\alpha}(\psi)$ and $E_{\alpha}(\tilde{\psi})$ are necessarily two different functions of $\alpha$ in the interval $[0,1]$, so that some $\alpha_0 \in [0,1]$ exists for which $E_{\alpha_0}(\psi) \neq E_{\alpha_0}(\tilde{\psi})$, say $E_{\alpha_0}(\psi) < E_{\alpha_0}(\tilde{\psi})$. Then monotonicity of $E_{\alpha_0}$ under LQCC forbids the local conversion $\psi \longrightarrow \tilde{\psi}$ to have efficiency one, since the process would imply an increase of $E_{\alpha_0}$. Therefore $\psi$ and $\tilde{\psi}$ are not locally equivalent.$\Box$

 We have shown therefore that, unless the two states are connected by local unitaries, at least one of the local transformations $\psi \longrightarrow \tilde{\psi}$ and $\tilde{\psi} \longrightarrow \psi$ can not be achieved with certainty \cite{probabilitat}. Later on in this section we will contrast this fact with well-established, asymptotic results. For this purpose let us now consider, in a ${\cal C}^3 \otimes {\cal C}^3$ system, the concrete conversion of the state $\psi$, with squared Schmidt coefficients $(0.5000,0.4991,0.0009)$, into the state $\tilde{\psi}$, with corresponding quantities $(0.7000,0.2737,0.0263)$. These two states have been chosen to have the same entropy of entanglement, 
\be
E_1(\psi) = E_1(\tilde{\psi}) = 1.0099 
\label{exemple1}
\ee
On the other hand we can use the rest of $\alpha$-entropies of entanglement and eq. (\ref{inequal}) to obtain upper bounds for the maximal probability $P(\psi\longrightarrow \tilde{\psi})$ of success in any local conversion. For instance, the upper bound coming from $E_{1/2}$ reads:
\be
P(\psi\longrightarrow \tilde{\psi}) \leq \frac{E_{1/2}(\psi)}{E_{1/2}(\tilde{\psi})} = 0.8754
\ee
 Notice that the same bound holds for the conversion of $N$ copies of $\psi$ into $N$ copies of $\tilde{\psi}$. Indeed, due to the additivity of the $\alpha$-entropies we find that $E_{1/2}(\psi^{\otimes N})= N E_{1/2}(\psi)$ for any $N$, and therefore
\be
P(\psi^{\otimes N}\longrightarrow \tilde{\psi}^{\otimes N}) \leq \frac{E_{1/2}(\psi^{\otimes N})}{E_{1/2}(\tilde{\psi}^{\otimes N})}=\frac{E_{1/2}(\psi)}{E_{1/2}(\tilde{\psi})}.  
\label{finit}
\ee
Finally, if we now refer to local strategies which convert the state $\psi^{\otimes N}$ into $\tilde{N}$ copies of $\tilde{\psi}$ with probability $p_{\tilde{N}}$, where some values $\tilde{N}$ might be a priori possible (cf. gambling in concentration of entanglement in \cite{BBPS}), we find that monotonicity of $E_{1/2}$ again sets an upper bound for the average number $<\!\tilde{N}\!> \equiv \sum_{\tilde N} p_{\tilde N} {\tilde N}$ of copies obtainable through any such strategies:
\be
<\!\tilde{N}\!>~ \leq ~ N \frac{E_{1/2}(\psi)}{E_{1/2}(\tilde{\psi})}.  
\label{exemple2}
\ee
 Consequently, for any finite $N$ there exists no strategy using only LQCC which transforms $N$ copies of $\psi$ into $N$ copies of $\tilde{\psi}$ with certainty, neither any local strategy which allows the parties to obtain, on average, $N$ copies of $\tilde{\psi}$ from $N$ copies of $\psi$, even though the entropy of entanglement of these two pure states is the same. Since similar results can be obtained for any two pure states we can announce

{\bf Statement:}
{\em The ratio $\frac{\tilde{N}}{N}$ of conversion of $N$ copies of a pure state $\psi$ into $\tilde{N}$ copies of another pure state $\tilde{\psi}$ is not given, typically, by the entropy of entanglement for any finite number of copies.}

\subsection{The asymptotic limit.}

 We have studied so far monotonicity under LQCC in the context of multipartite systems which are described by a finite dimensional Hilbert space. We move now to consider a well studied limiting case, involving pure states of the form $\psi^{\otimes N}$ for $N$ arbitrarily large.

 Bennett et al. \cite{BBPS} showed that $N$ copies of any pure state $\psi$ can be {\em asymptotically} converted into $N\frac{E_1(\tilde{\psi})}{E_1(\psi)}$ copies of any other pure state $\tilde{\psi}$ using only LQCC. Three features characterize the transformations involved in such conversions \cite{composicio}. Firstly, for any finite $N$ the transformation is {\em approximate}: its outcome $\xi$ only resembles the wished final state $\tilde{\psi}^{\otimes \tilde{N}}$. A physically sensible measure of the degree of approximation in the transformation is then given by the fidelity, which is required to be
\be
F(\xi,\tilde{\psi}^{\otimes \tilde{N}}) \geq 1-\epsilon,
\ee
where $\epsilon > 0$ for any finite $N$. Secondly, there is a priori {\em no certainty} in the transformation being successful. That is, the probability of success $P_{\epsilon}(\psi^{\otimes N} \longrightarrow\tilde{\psi}^{\otimes \tilde{N}}) \equiv P(\psi^{\otimes N} \longrightarrow \xi)$ in the approximate conversion is typically smaller than one, and bounded through a second parameter $\eta > 0$ by
\be
P_{\epsilon}(\psi^{\otimes N} \longrightarrow\tilde{\psi}^{\otimes \tilde{N}}) \geq 1-\eta.
\ee
And thirdly, the number $\tilde{N}$ of copies of $\tilde{\psi}$ the parties end up sharing is not $N\frac{E_1(\tilde{\psi})}{E_1(\psi)}$ but $N\frac{E_1(\tilde{\psi})}{E_1(\psi)}(1-\delta)$, $\delta > 0$, for any finite $N$. Then, the asymptotic results in \cite{BBPS} can be rewritten to state that for any required $\epsilon, \eta, \delta > 0$ a sufficiently large $M$ always exists such that, for any $N \geq M$, the conversion of $N$ copies of $\psi$ into a $\xi$ very close to $N\frac{E_1(\tilde{\psi})}{E_1(\psi)}(1-\delta)$ copies of $\tilde{\psi}$ is possible with probability greater than or equal to $1 -\eta$, that is $\forall \epsilon, \eta, \delta > 0 ~~\exists M/~~ \forall N \geq M,$
\be
P_{\epsilon}(\psi^{\otimes N} \longrightarrow\tilde{\psi}^{\otimes N\frac{E_1(\tilde{\psi})}{E_1(\psi)}(1-\delta)}) \geq 1-\eta,
\ee
or equivalently,
\be
\lim_{\epsilon, \delta \rightarrow 0}\lim_{N\rightarrow\infty} P_{\epsilon}(\psi^{\otimes N} \longrightarrow\tilde{\psi}^{\otimes N\frac{E_1(\tilde{\psi})}{E_1(\psi)}(1-\delta)}) = 1.
\label{limit}
\ee
The asymptotic ratio of conversion reveals that in this limiting case entanglement transformations between pure states can be made reversible, and that the entropy of entanglement $E_1$, which gives the optimal ratio achievable in an asymptotic conversion, quantifies uniquely \cite{PopRo} entanglement in this regime. These results are of fundamental interest in quantum information theory, and we would like to analyze next what consequences they have for EMs. 

 It follows from eq. (\ref{limit}) that the entropy of entanglement is the only EM for pure states in the asymptotic limit \cite{PopRo}. Indeed, any other magnitude which behaves monotonically under LQCC in the finite regime can be understood to lose its monotonicity in the limit $N \rightarrow \infty$, provided that allowances are made for some inefficiency and approximation in the conversions for any finite $N$. We find it instructive to report what happens to the $\alpha$-entropies of entanglement for $\alpha < 1$ as $N$ grows, for it enlightens about the fate of EMs (which are not the entropy of entanglement) in the asymptotic regime. For this purpose we have considered in the appendix the protocol of entanglement dilution that was presented in \cite{BBPS}, and that is asymptotically optimal. Since $E_{\alpha}$ are additive magnitudes, they necessarily diverge when evaluated on $\psi^{\otimes N}$ for $N\rightarrow \infty$, whereas the intensive measures $e_{\alpha} \equiv \frac{1}{N}E_{\alpha}$ remain finite and become suitable for our analysis. It follows from the appendix that as N grows any $e_{\alpha}$ tends to a discontinuous function of the fidelity for $\alpha < 1$ (cf. eqs. (\ref{exemplea})-(\ref{exempleb})). Therefore, for large enough $N$, two states $\xi$ and $\psi^{\otimes N}$ may be very similar physically (i.e. $F(\xi,\psi^{\otimes N})$ may be very close to one) and yet their corresponding $e_{\alpha}$ ($\alpha < 1$) may differ significantly. More specifically, we show in the appendix that a collection of states $\xi_N$ exist such that, for $\alpha < 1$, $lim_{N\rightarrow \infty} F(\xi_N,\psi^{\otimes N}) = 1$ and yet $0=lim_{N\rightarrow \infty} e_{\alpha}(\xi_N) \neq lim_{N\rightarrow \infty} e_{\alpha}(\psi^{\otimes N}) = E_{\alpha}(\psi)$. This makes $e_{\alpha < 1}$ unsuitable for quantifying the non-local resources of $\xi_N$ and $\psi^{\otimes N}$, $N\rightarrow\infty$, since such measures make distinction between these two states, which are physically equivalent ---and in particular can be used to accomplish the same tasks requiring entanglement. 

Notice however that, as the example in eqs. (\ref{exemple1}-\ref{exemple2}) shows, the limits in eq. (\ref{limit}) do no commute. That is if one requires certainty ($\eta = 0$) in the conversion and does not allow for it to be approximate ($\epsilon = 0$), then the bound in eq. (\ref{exemple2}) holds for any $N$. This indicates that uniqueness of EMs in the asymptotic limit is compatible with having many of these quantities in the finite regime. Nevertheless, since small values for $\eta$ and $\epsilon$ are most often acceptable, for two states whose fidelity is very close to one are almost indistinguishable (their non-local resources are practically the same), and a conversion with almost certainty of success almost always does the job, the entropy of entanglement is singled out as a privileged measure of the non-local resources also for a finite, sufficiently large number $N$ of copies of a pure state, if one allows for small $\epsilon,\eta,\delta > 0$, as it is most often reasonable in realistic situations.

\section{Measures of entanglement}
\label{seccio5}
 
 Quantifying entanglement is a priority in quantum information theory, for many of its applications rely on quantum correlations as a necessary resource. The ultimate purpose of this work is to contribute to a better understanding of how entanglement should be quantified. We propose here monotonicity under LQCC as the only requirement any reasonable measure of entanglement has to satisfy \cite{condicions}. It will be argued also that only one measure is not sufficient to completely quantify entanglement of pure states for bipartite systems, while suggesting that, on the contrary, several independent measures should be employed simultaneously. 
 
\subsection{Entanglement monotones as measures of non-local resources.}

 Let us discuss first our point of view about the properties one should require from a measure of entanglement. We understand that from a proper quantification of the non-local resources of a shared state the parties should be able to infer what jobs they can carry out with them and to what extend. It should inform also of what non-local resources are needed to prepare the state. Notice that when quantified in this terms, entanglement has a monotone behavior under local transformations. Indeed, by mere tautology, the extend to which the parties can accomplish a given task when sharing a state can not increase, on average, under LQCC, for LQCC determine precisely the transformations the parties are allowed to perform to accomplish the task. Similarly, the cost of preparation ---by means of LQCC and in terms of other states--- of the state of a system is a magnitude that does not increase under local transformations. Consequently, monotonicity under LQCC is a natural requirement for any magnitude aiming at quantifying entanglement when regarded as a resource.

 Conditions C1 and C2, characterizing monotonicity under LQCC (eqs. (\ref{C1}) and (\ref{C2})), admit then a straightforward reinterpretation as requirements for a measure of entanglement. Condition C1 (to be compared with those in \cite{VPRK}, see also \cite{VedrPl,PopRo}) simply states that the parties can not make the non-local resources grow systematically by operating locally on the system. Notice that this condition alone suffices to characterize monotonicity under local transformations between pure states: remarkably enough monotonicity under LQs implies automatically monotonicity under LQCC for multipartite pure states. On its turn condition C2 states that by dismissing information about the shared system, the parties will not be able to make a better use of its resources than previously to the dismissal, nor to make the state become of more expensive preparation (again, as above, by definition). Finally, recall that classical correlations can be introduced between the subsystems by means of correlated local operations. Then condition C2 is necessarily satisfied by any measure which does not mistake these classical correlation for genuinely quantum ones.

\subsection{Minimal set of entanglement monotones.} 

 The entanglement of bipartite pure states is completely quantified by a unique measure $\cite{PopRo}$, the entropy of entanglement, in the asymptotic limit. We have shown in this work that one magnitude does not suffice in general to characterize the non-local resources of $\psi \in {\cal C}^n \otimes {\cal C}^n$ for it takes $n-1$ independent Schmidt coefficients to determine its class of equivalence under LQCC. We have also shown, by means of their explicit construction, that infinitely many EMs exist, and now propose all of them as candidates for measures of entanglement. We envisage that the quantification of the non-local resources will be possible by means of a {\em set} of measures. It would be then interesting to find, among all EMs, a {\em minimal set} of them which would quantify entanglement completely. In this direction we have shown that at least $n-1$ independent measures are to be employed for bipartite pure states $\psi \in {\cal C}^n \otimes {\cal C}^n$.

\section{Concluding remarks}

 There is still a lot of progress to be made concerning the quantification of non-local resources. At the core of the problem we find the set of local transformations, which exhaust all possible changes the state of a multipartite system can experience when the parties sharing it act restricted to LQCC. The quantification of entanglement as a resource is to be understood in relation to these transformations.

 We have focused in this work on those magnitudes which behave monotonically under local transformations. We have given a mathematical characterization of such monotones in terms of elementary local transformations and have shown how they can be used to determine the optimality of local conversions of shared states. A more detailed study for the case of pure states of bipartite systems, for which we have shown how any monotone can be constructed, has revealed the need for several independent measures of entanglement to be used simultaneously. We have also shown that this is not in contradiction with the asymptotic uniqueness of measures for pure states.  Posterior results have confirmed the validity of our approach \cite{Nie,Vid,Har,JonPl}. 

 It would be interesting to obtain similar results for pure states shared by three or more parties. The fact that no trivial generalization of the Schmidt decomposition is known to exist in this case means a serious difficulty. A deeper study of monotonicity for mixed states is also challenging, as quantum correlations coexist with classical ones in a mixed state. 

{\bf Acknowledgments}

 The author would like to thank Rolf Tarrach for thoroughly reading the manuscript, also for many comments, suggestions and corrections, which have certainly helped to improve this paper. The author is also grateful to Michal  Horodecki, Vlatko Vedral and Martin B. Plenio for suggestive discussions, and wishes to thank the participants of the Benasque Center for Physics '98 session for comments and criticism. Financial support from CIRYT, contract AEN98-0431, and CIRIT, contract 1998SGR-00026, and a CIRIT grant 1997FI-00068 PG are also acknowledged.

\appendix
\section{Asymptotic behavior of the $\alpha$-entropies of entanglement.}

 Let us first remind how the dilution of entanglement can be performed with asymptotically maximal efficiency. The protocol described in \cite{BBPS} uses both quantum data compression and quantum teleportation. Alice and Bob share initially $N$ states $\ket{\psi} \equiv 1/\sqrt{2}(\ket{1_A\otimes 1_B} + \ket{2_A\otimes 2_B}) \in {\cal C}^2 \otimes{\cal C}^2$, with $E_1(\psi) = 1$. Then Alice prepares locally $\tilde{N}$ copies of the state $\ket{\tilde{\psi}_{AA'}} \equiv \cos\theta \,\ket{1_A\otimes 1_{A'}} + \sin\theta \,\ket{2_A\otimes 2_{A'}}$ for some $\theta \in (0,\frac{\pi}{4})$, i.e. the state $\tilde{\psi}^{\otimes \tilde{N}}_{AA'}$, whose Schmidt decomposition has $2^{\tilde N}$ coefficients. These coefficients can be grouped into $\tilde{N}+1$ sets, the $l$-th set containing $(\begin{array}{c}\!\tilde{N}\! \\\! l\!\end{array})$ identical coefficients of the form $\cos^{\tilde{N}-l}\theta\sin^l\theta$ ($l=0,1,...,\tilde{N}$). Now, by employing $\tilde{N}$ copies of $\psi$ to teleport, Alice and Bob can end up sharing the state $\tilde{\psi}^{\otimes \tilde{N}} \equiv \tilde{\psi}^{\otimes \tilde{N}}_{AB}$, and this is not possible deterministically if less than $\tilde{N}$ copies of $\psi$ are used. Nevertheless by teleporting with less than $\tilde{N}$ copies, say with just a number $M(r)\equiv \log_2 \sum_{l=0}^r (\begin{array}{c}\!\! \tilde{N}\!\\ l\!\end{array})$ of them, they can end up sharing a state which is, essentially, a projection $\xi_{\tilde{N}}(\frac{r}{\tilde{N}})$ of $\tilde{\psi}^{\otimes \tilde{N}}$ containing its greater $\sum_{l=0}^r (\begin{array}{c}\!{\tilde  N}\! \\\! l\!\end{array})$ Schmidt coefficients and that may well do the job if one allows the transformation to be approximate. That is, instead of the transformation
\be
\psi^{\otimes \tilde{N}} \longrightarrow \tilde{\psi}^{\otimes \tilde{N}},
\ee
the parties may consider performing alternatively the transformation
\be
\psi^{\otimes M(r)} \longrightarrow \xi_{\tilde{N}}(\frac{r}{\tilde{N}}),
\ee
where $M(r) < \tilde{N}$.

How similar is the projection $\xi_{\tilde{N}}(\frac{r}{\tilde{N}})$ to the desired state $\tilde{\psi}^{\otimes \tilde{N}}$ as a function of the fraction $x\equiv r/\tilde{N}$? The computation of its fidelity gives
\begin{equation}
F_{\tilde N}(x) \equiv |\braket{\xi_{\tilde{N}}(x)}{\psi^{\otimes \tilde{N}}}|^2 = (\sum_{l=0}^{x\tilde{N}} (\begin{array}{c}\!\tilde{N}\! \\\! l\!\end{array})\cos^{2(\tilde{N}-l)}\theta\sin^{2l}\theta)^2.
\end{equation}
Thus as $\tilde{N}$ grows the fidelity $F_{\tilde{N}}(x)$ tends to the step function $\Theta(x-x^*)$, for a fraction $x^*$ such that $M(\tilde{N}x^*)=\tilde{N}E_1(\tilde{\psi})$ (i.e. $x^*$ corresponds to using $\tilde{N}E_1(\tilde{\psi})$ copies of the initially shared state $\psi$ to teleport). On the other hand the entropy of the entanglement of $\xi_{\tilde{N}}(x)$,
\begin{equation}
E_{1,\tilde{N}}(x) \equiv \sum_{l=0}^r (\begin{array}{c}\!\tilde{N}\! \\\! l\!\end{array})\cos^{2(\tilde{N}-l)}\theta\sin^{2l}\theta \log_2 \cos^{2(\tilde{N}-l)}\theta\sin^{2l}\theta,
\end{equation}
behaves similarly as $\tilde{N}$ grows, since
\begin{equation}
\lim_{\tilde{N} \rightarrow \infty} \frac{1}{\tilde{N}} E_{1,\tilde{N}}(x) = E_1(\tilde{\psi})\Theta(x-x^*).
\end{equation}

 Now, what does happen, as $\tilde{N}$ is increased, to the rest of $\alpha$-entropies of entanglement? Take $E_{\alpha}$ for any fixed $\alpha <1$. A direct computation of $E_{\alpha}(\xi_{\tilde{N}}(x))$ shows that, for any $\beta > 0$, a $\delta(\alpha,\beta)>0$ and an integer $M(\alpha,\beta)$ exist such that for any $\tilde{N}\geq M$ the $\alpha$-entropy of entanglement of the projection $\xi_{\tilde{N}}(x^*+\delta)$ is smaller than $\tilde{N}\beta$. This implies that for any $E_{\alpha}$ which is not the entropy of entanglement a projection $\xi_{\tilde{N}}(x^*+\delta)$ exists such that
\begin{equation}
\lim_{\tilde{N} \rightarrow \infty} \frac{1}{\tilde{N}} E_{\alpha}(\xi_{\tilde{N}}(x^*+\delta)) = 0,
\end{equation}
but for which, on the other hand,
\begin{equation}
\lim_{\tilde{N} \rightarrow \infty} \frac{1}{\tilde{N}E(\tilde{\psi})} E_{1,\tilde{N}}(x^*+\delta(\alpha)) = \lim_{\tilde{N} \rightarrow \infty} F_{\tilde{N}}(x^*+\delta(\alpha)) = 1.
\end{equation}
 This shows, when considering $\frac{1}{\tilde{N}} E_{\alpha}(\xi_{\tilde{N}}(x))$ as a function of $F_{\tilde{N}}(x)$, i.e. $\frac{E_{\alpha}}{\tilde{N}}(F_{\tilde{N}})$, that it tends indeed to a discontinuous function of the fidelity in the limit $\tilde{N} \rightarrow \infty$ for $\alpha < 1$.


\begin{references}

\bibitem{BBPS} C.H. Bennett, H.J. Bernstein, S. Popescu, B. Schumacher, Phys. Rev. A {\bf 53} 2046 (1996).

\bibitem{BdVSW} C.H. Bennett, D.P. DiVincenzo, J.A. Smolin, W.K. Wootters, Phys. Rev. A {\bf 54} 3824 (1996).

\bibitem{VedrPl} V.Vedral, M.B. Plenio, Phys. Rev. A {\bf 57}, 1619 (1996).

\bibitem{LoPop} H.-K. Lo and S. Popescu, "Concentrating entanglement by local actions-beyond mean values", quant-ph/9707038.

\bibitem{Nie} M.A. Nielsen, "A partial order on the entangled states", quant-ph/9811053.

\bibitem{Vid} G. Vidal, "Entanglement of pure states for a single copy", quant-ph/9902033. 

\bibitem{Har} L. Hardy, "A method of areas for manipulating the entanglement properties of one copy of a two-particle pure state", quant-ph/9903001.

\bibitem{JonPl} D. Jonathan and M.B. Plenio, "Minimal conditions for local pure-state entanglement manipulation", quant-ph/9903054. 

\bibitem{KLM} A. Kent, N. Linden and S. Massar, "Optimal entanglement enhancement for mixed states", quant-ph/9902022 
 

\bibitem{PopRo}  S. Popescu, D. Rohrlich, Phys. Rev. A {\bf 56} R3319 (1997).

\bibitem{VPRK} V. Vedral, M.B. Plenio, M. A. Rippin and P.L. Knight, Phys. Rev. Lett. {\bf 78}, 2275 (1997).

\bibitem{VT} G. Vidal, R. Tarrach, "Robustness of entanglement", quant-ph/9806094

\bibitem{finit} Notice that the asymptotic scenario, to which we referred in the introduction and to which we will come back in sec. \ref{seccio4}, requires a global Hilbert space of infinite dimension.

\bibitem{Kraus} K. Kraus, States, Effects and Operations: Fundamental Notions of Quantum Theory, Springer-Verlag, Berlin, 1983. See, for a review, also M.A. Nielsen, C.M. Caves, B. Schumacher and H. Barnum, "Information-theoretic approach to quantum error correction and reversible measurement", quant-ph/9706064. 

\bibitem{canvi} The local Hilbert space of party A, ${\cal H}_A$, is replaced with ${\cal H}_A'$ if the local operation ${\cal E}_{A,k}$ involves that party A throw away a part of its subsystem or append ancillas to it. Thus $A_{k,j}$ is in general an operator from ${\cal H}_A$ to ${\cal H}_A'$.

\bibitem{external} Notice that when CC between the parties is allowed there is no need for an external observer in order for the parties to be able to have an equivalent description of the system.

\bibitem{explanation} Thus both $\rho_1$ and $\rho_2$ are density matrices acting in a Hilbert space ${\cal H}_A' \otimes {\cal H}_B'$, so that the Hilbert space of the two-particle subsystem held by Alice (Bob) is ${\cal H}_A = {\cal H}_A'^{\otimes 2}$ (${\cal H}_B = {\cal H}_B'^{\otimes 2}$).

\bibitem{separable} We call separable any state which can be written as a convex combination $\sum_k p_k \rho_{A,k}\otimes\rho_{B,k}\otimes ...$ of states which are there direct product of states $\rho_{A,k}, \rho_{B,k}, ...$ describing the local subsystem of {\em each} party. They correspond to the {\em fully separable states} as described in W.Dur, J.I. Cirac and R. Tarrach, "Separability and distillability of multiparticle quantum systems", quant-ph/9903018.

\bibitem{exemple} Suppose, for instance, that a black-box-like device ${\cal A}$ on Alice's side dismisses one of Alice's particles according to an internal random mechanism, and that its analog ${\cal B}$ on Bob's side receives then a classical signal from ${\cal A}$, which allows it to throw away the particle corresponding to the same pair. 

\bibitem{info} 
Information in local transformations concerns the initial state of the system and the operations being performed (including the actual performance in a stochastic transformation or the actual outcome of a measurement). In a given transformation some of this information may be either {\em available} or not  (e.g. in transformation (\ref{ambinfo}) the parties know which pair is actually kept, while they ignore it in transformation (\ref{senseinfo})). We have adopted throughout this work the following point of view regarding the information about the multipartite system: the parties, allowed to communicate classically, always share all information which is available to any of them. In addition, this information is acquired as soon as it is produced (i.e. if, e.g., the result of a measurement is to be known at some point, then we consider that it is communicated to all parties as soon as the measurement is performed). Thus, the only way new information can be acquired is through direct local interaction with the system, whereas any loss of information is irreversible.

\bibitem{llista} Examples of entanglement monotones are the {\em entanglement of formation} and {\em distillable entanglement} \cite{BdVSW}, {\em relative entropy of entanglement} \cite{VedrPl} and the {\em robustness of entanglement} \cite{VT}.

\bibitem{probabilitat} The maximal probability of success in the  $\psi\longrightarrow\tilde{\psi}$ has been presented in a posterior work \cite{Vid}. 

\bibitem{simplex} The set of vectors $\vec{\alpha}$ of possible eigenvalues of a density matrix $\sigma$ acting on ${\cal C}^n$ is the simplex $\Delta_n$, defined by the constrains $\alpha_i \geq 0$, $\sum_{i=1}^n \alpha_i = 1$. That $g$ in eq. (\ref{funciog}) has to be concave follows from considering eq. (\ref{A2}) when $\sigma_1$ and $\sigma_2$ commute. That $g$ has to be symmetric under exchange of arguments from eq.(\ref{A1}). Notice that it might not be true that any symmetric, concave function $g$ be related via eq. (\ref{funciog}) to a unitarily invariant concave function $f$, since concavity of $g$ does not imply that of $f$.

\bibitem{We} A. Wehrl, Rev.Mod.Phys. {\bf 50} 221 (1978)

\bibitem{Renyi} A. Renyi, Probability Theory, North-Holland, Amsterdam, 1970.

\bibitem{reversible} We have learned of a posterior work by Nielsen \cite{Nie}, where he presents the conditions for one state $\psi$ to be convertible, with certainty, into another state $\tilde{\psi}$ using only LQCC. His (more powerful) results lead to the same conclusion as far as local equivalence of pure states is concerned. 

\bibitem{condicions} Conditions C1 is to be compared with conditions presented in \cite{VPRK}. See also \cite{VedrPl} and \cite{PopRo}. Condition C2 was presented in \cite{VT}. Let us remark that we do not demand additivity for measures of entanglement, whereas some other authors consider it a natural requirement (see, for instance, A.V. Thapliyal, "On multipartite pure-state entanglement", quant-ph/9811091).

\bibitem{composicio} Transformations which asymptotically reach the optimal ratio of conversion dictated by the entropy of entanglement can be obtained, for a generic conversion, by composing suitable generalizations of the strategies of entanglement concentration and dilution presented in \cite{BBPS}.

\end{references}
\end{document}